\documentclass[aps,twocolumn,prd,nofootinbib,showpacs]{revtex4}
\usepackage[pdftex,bookmarks]{hyperref}

\usepackage[T1]{fontenc}
\usepackage{amsmath,amssymb}

\usepackage{graphics,wrapfig,times}
\usepackage{graphicx}
\usepackage{color}
\usepackage{mathrsfs}

 \makeatletter
 \newcommand{\pback}[1]{{
   \let\@rrow=\leftarrowfill
   \mathchoice{\AIN@stemPullBack{#1}{\@rrow}}{\AIN@stemPullBack{#1}{\@rrow}}
     {\AIN@indxPullBack{#1}{\@rrow}}{\AIN@indxPullBack{#1}{\@rrow}}}
   \vphantom{#1}}

 \newcommand{\AIN@stemPullBack}[2]{
   \vtop{\mathsurround=0pt
   \ialign{##\crcr$\textstyle{#1}\strut$\crcr
     \noalign{\kern-0.4ex\nointerlineskip}{\tiny#2}\crcr}}}

 \newcommand{\AIN@indxPullBack}[2]{
   \vtop{\mathsurround=0pt
   \ialign{##\crcr\hfil$\scriptstyle{#1}$\hfil\crcr
     \noalign{\kern+0.4ex\nointerlineskip}{\tiny#2}\crcr}}}

\def\bar{\overline}
\def\be{\begin{equation}}
\def\ee{\end{equation}}
\def\bea{\begin{eqnarray}}
\def\eea{\end{eqnarray}}
\def\ba{\begin{array}}
\def\ea{\end{array}}
\def\nn{\nonumber}
\def\w{\wedge}

\def\a{\alpha}
\def\b{\beta}
\def\c{\gamma}
\def\d{\textrm{d}}

\definecolor{D}{rgb}{0.00,0.17,0.48}
\definecolor{M}{rgb}{0.00,0.02,0.83}
\definecolor{L}{rgb}{0.58,0.79,1.00}
\definecolor{R}{rgb}{0.80,0.00,0.00}
\definecolor{G}{rgb}{0.02,0.40,0.10}

\begin{document}

\title{Local and global structure of domain wall space-times}
\author{ Yu-Huei Wu$^{a,c}$} \email{yhwu@phy.ncu.edu.tw}
\author{Chih-Hung Wang$^{b,c}$} \email{chwang@phy.ncu.edu.tw}

\affiliation{$^{a}$ Center for Mathematics and Theoretical Physics, National Central University, Chungli 320, Taiwan, R.O.C.\\
$^{b}$ Department of Physics, Tamkang University, Taipei 25137, Taiwan, R.O.C.\\
$^{c}$ Department of Physics, National Central University, Chungli 320, Taiwan, R.O.C.}


%
\begin{abstract}

  We present a general proof on the equivalence of the comoving-coordinate approach, where the wall is fixed at a constant coordinate variable, and moving-wall approach, where the wall is moving in a background static space-time, in the domain wall space-times without reflection symmetry. We further provide a general procedure to construct the comoving coordinates in the domain wall space-times, where the two regions separated by an infinite thin wall have different cosmological constant $\Lambda$ and Schwartzschild mass $M$. By solving Israel's junction conditions in the thin-wall limit, the gravitational fields of spherical, planar and hyperbolic domain wall space-times with $M=0$ in the two different comoving coordinate systems are obtained. We finally discuss the global structure of these domain wall space-times.

\end{abstract}
\date{\today}
\maketitle

\section{Introduction}


It is generally believed that phase transitions are occurred in the early Universe, so various types of topological defects can naturally form by Kibble mechanism \cite{Kibble-76} (see \cite{VS} for a review).  Domain walls, a particular type of the topological defects, correspond to vacuumlike hypersurfaces interpolating between separate vacua. Beside the Kibble mechanism, domain walls can also form as the boundary of a true vacuum bubble created by quantum tunneling process of false vacuum decay, i.e. bubble nucleation \cite{CD-80, Coleman-85}, and the dynamics of bubbles has been studied in the framework of general relativity \cite{BKT-87, AKMS-87}. In the study of gravitational effects of thin walls or dynamics of vacuum bubbles, it is useful to apply the thin-wall approximation, which is also considered in this paper.

In the thin-wall approximation, the wall is regarded as an infinitely thin with $\delta$-function singularity in the energy-momentum tensor. The two regions $V^+$ and $V^-$ separated by the wall may have different physical parameters, e.g. cosmological constant \{$\Lambda_{+}, \Lambda_{-}$\} and Schwartzschild mass \{$M_{+}, M_{-}$\}. In this paper, we denote $V^{+}(V^{-})$ for exterior (interior) region to the wall and any quantity $Q$ with a subscript $+ (-)$ corresponds to the quantity at $V^{+}(V^{-})$.  Therefore gravitational effects of domain walls are described by Einstein's field equations off the wall together with Israel's junction conditions \cite{Israel-66, IS-84}. As far as we know,  domain wall solutions have been studied based on two different approaches. The first approach, which we call moving-wall (MW) approach,  starts from exact solutions of Einstein's field equations off the wall in the specific coordinates, e.g. the Schwartzschild coordinates, and then the wall's motion in the specific background metric is obtained by satisfying Israel junction conditions.  The second approach, i.e. comoving-coordiantes (CC) approach, is to introduce the co-moving coordinates, where the wall is placed at a particular constant coordinate variable, say $z=z_0$, and domain wall solutions are obtained by solving Einstein's field equations off the wall and Israel's Junction conditions.

Since domain wall solutions obtained from these two approaches both satisfying Einstein's field equations and Israel's junction condition, one may expect that these two different approaches are equivalent up to a coordinate transformation. In the study of brane cosmologies, Mukohyama {\it et al} \cite{MSM-00} has shown the coordinate transformation between two exact solutions of brane world, which are obtained by CC approach in Gaussian normal coordinates \cite{BDEL-00, M-00} and MW approach \cite{Ida-00}, respectively.   Moreover, Bowcock \textit{et al} \cite{BCG-00} demonstrated the equivalence of these two approaches in brane-cosmological models with $Z_2$ symmetry\footnote{In \cite{BCG-00}, these two different approaches are called the brane-based approach and the bulk-based approach.}, and then studied time evolution of 4-dimensional brane-universe in the MW approach without $Z_2$ symmetry. However, it is still not clear to us how to construct a comoving coordinates in domain wall space-times without $Z_2$ symmetry by given a trajectory of wall's motion in the MW approach.  In this paper, we present a general proof on the equivalence of these two approaches in the 4-dimensional spherical, planar, and hyperbolical domain wall space-times without $Z_2$ symmetry, and then show how to construct comoving coordinates in domain wall space-times having different $\Lambda$ and $M$ on each side of the wall.

Although Einstein's field equations and also field equations of any classical field are general covariant according to general principle of relativity, finding a proper coordinate system to describe the dynamics of classical quantities and gravitational fields are still important, especially when one wants to compare the results to observations. For example, in the post-Newtonian approximation \cite{Weinberg, MTW, Will-93},  equations of motion of many bodies are described in a particular coordinate system, where one can obtain Newtonian theory of gravity in the first-oder approximation of Einstein's field equations. Since the dynamics of our solar system are well described in Newtonian gravity, it indicates that this coordinate system is a proper choice to study our solar system. So higher-order effects of general relativity in this coordinate system can be calculated and compared to solar system observations.  Furthermore, in the study of quantum fields in curved space-time, the coordinate choices become significant since there is no coordinate-invariant definition of the vacuum state, i.e. the vacuum state is coordinate-dependent \cite{BD-82}. Besides showing the construction of comoving coordinates in domain wall space-times, our another motivation is to construct  a proper comoving coordinate system in spherical, planar, and hyperbolic domain wall space-times, which may be useful to understand gravitational effects of domain walls on primordial quantum fluctuations during inflation.

 From the generalized Birkhoff theorem \cite{Goenner-70, GS-70, KSHM-80},  the exact solutions of Einstein's field equation with cosmological constant $\Lambda$ in spherically, planar, and hyperbolically symmetric space-times can be written as
\bea
g= -U(q) \,\d T \otimes \d T + \frac{1}{U(q)} \d q \otimes \d q + q^2  \d V_2, \label{static-g}
\eea where $U(q)=k- 2M/q - (\Lambda/3) q^2$, $\d V_2 = (1-k x^2)^{-1} \d x \otimes \d x +x^2 \d \phi \otimes \d \phi$, and $k$ denotes constant Gaussian curvature. It is clear that the metric (\ref{static-g}) is static in the certain range of coordinates. In the MW approach, the metric $g_{+}\,(g_{-})$ in $V^{+}\, (V^{-})$ is described by $U_{+(-)} $, where $U_{+(-)} = k- 2M_{+(-)} /q - (\Lambda_{+(-)}/3) q^2$.\footnote{To concise our notation, we use subscript $\pm$ on any quantity $Q$ to denote $Q_{+}(Q_{-})$ in this paper.} Thus the Israel's junction conditions yield equations of motion of domain walls, which have been studied in \cite{BKT-87, AKMS-87}, and the wall's trajectories are described in terms of proper-time $\tau$.  To verify the equivalence of MW and CC approaches, we consider the wall being placed at $r=r_0$, and by using Einstein's field equations, metric continuity (with requiring coordinate time $\eta$ on the wall being $\tau$), and Israel's junction conditions, we derive that metric at $r=r_0$, which is denoted by $\hat{g}$, satisfies the same equations as equations of motion of the wall obtained in the MW approach \cite{AKMS-87, BKT-87}. We further find that $\hat{g}$ will uniquely determine the metric in $V^{+}$ and $V^{-}$. It means that once the $\hat{g}$ is known, we can then obtain $g_+$ and $g_-$.




Since there exists a degree of freedom on $\hat{g}$ due to the choice of the time coordinate on the wall, we calculate $g_{+}$ and $g_{-}$ in the two different comoving coordinate systems by requiring $\hat{g}_{00}=-1$ (case I) and $\hat{g}_{00}=-\alpha^2/\eta^2$ (case II), where $\hat{g}_{00}$ denotes the metric component $g_{00}$ on the wall and $\alpha$ is a constant. So case I indicates that the coordinate time on the wall is the proper-time. Interestingly, we find that the  metric solutions $g_{\pm}$ in case I with $M{\pm}=0$ are the same as the domain wall solutions obtained by Cveti\v{c} {\it et al} \cite{CGS-93}. In \cite{CGS-93}, they obtained domain wall solutions in the comoving coordiantes by using a metric ansatz:
\bea
g= A(z) (- \d t \otimes \d t + \d z \otimes \d z + S^2 (t)\,\d V_2).
\eea However, we start from a general metric form in spherical, planar, and hyperbolic symmetric space-time and it turns out that the domain wall solutions in case I with $M{\pm}=0$ agree with \cite{CGS-93}. In our previous work \cite{WCW-11}, we obtained a planar domain wall solution with reflection symmetry in de Sitter space-time, and $\sigma_0=0$ yields the well known metric of steady-state Universe in the conformal time coordinate. We further study its gravitational effects on primordial quantum fluctuations during inflation, and found that its gravitational fields produce a primordial dipole effects in the power spectrum of primordial curvature perturbation \cite{WWH-12}. It is observed that this planar domain wall solution does satisfy $\hat{g}_{00}=-\alpha^2/\eta^2$ and the coordinate time $\eta$ on the wall corresponds to conformal time in de Sitter spacetime, we suggest that the choice of  $\hat{g}_{00}=-\alpha^2/\eta^2$ may provide a proper coordinates to investigate the gravitational effects of domain walls in the early Universe.
Since our previous domain wall solution requires plane and reflection symmetry, it is quite limited to study of gravitational fields of realistic domain wall space-times. For example, false vacuum decay yields two space-time regions with different $\Lambda$ separated by a spherical bubble. In the study of case II, we generalize our previous planar domain wall solution to spherical, planar, and hyperbolic domain wall space-times without reflection symmetry.

The global structure of spherical, planar and hyperbolic domain wall space-times has been well studied in \cite{CGS-93, BKT-87} (see a review article \cite{CS-97} and the references in). Moreover, Ref. \cite{CS-97} has pointed out that the constant-$r$ sections of non- and ultraextreme domain wall space-times, which correspond to domain wall solutions in  case I with $H^2\neq0$,  all represent (2+1)-dimensional de Sitter space-time ($\d\textrm{S}_3$), whose topology is $\textbf{R}$ ({time}) $\times$ $\textbf{S}^2$ ({space}). It turns out that non- and ultraextreme planar domain wall, which is locally plane-symmetric and geodesically incomplete, describes only a part of a spherical bubble \cite{CS-97}. In Sec. \ref{4}, we present a coordinate transformation between the domain wall solutions in case I and solutions in case II, so topology of spherical, planar, and hyperbolic domain walls in case II is also  $\textbf{R}$ ({time}) $\times$ $\textbf{S}^2$ ({space}). Hence, one should expect that the planar domain wall solution in case II also represents a portion of a spherical domain wall space-time.

The plan of this paper is as follows. In Sec. \ref{2}, we show the equivalence of CC and MW approaches and also the construction of comoving coordinates in domain wall space-time with different $M_{\pm}$ and $\Lambda_{\pm}$.  Sec. \ref{3} discuss spherical, planar, and hyperbolic domain wall solutions in two different comoving coordinates, i.e. case I and case II, with $M_{\pm}=0$. Moreover, since the choice of $\hat{g}_{00}=-\alpha^2/\eta^2$ is motivated to study gravitational effects of domain walls on quantum fluctuations during inflation, we will only discuss $\Lambda_{\pm}>0$ in case II.  In Sec. \ref{4}, the global structure of domain wall space-times are discussed. Sec. \ref{5} gives a discussion and conclusion. In Appendix \ref{A}, we present some technical materials, which is useful for following our calculations in Sec. \ref{2} and Sec. \ref{3}.


We use the units $\hbar=c=1$, and the metric signature is $(- + + +)$. The Latin indices $a, b, \cdots$ are referred to coordinate indices and the Greek indices $\a,\b,\gamma \cdots$ referred to orthonormal frame indices. $g$ and $\nabla$ denote metric tensor and Levi-Civita connection, respectively.

\section{On the equivalence of comoving-coordinate and moving-wall approaches} \label{2}

In the thin-wall approximation, the thickness $\varepsilon$ of a thin wall is taken to be zero, so the infinitely thin wall becomes a 3-dimensional timelike, null or spacelike hypersurface $\Sigma$ in  4-dimensional space-times, and its associated stress-energy tensor $T^a{_b}$ of the space-times has a $\delta$-function singularity on $\Sigma$. Here, we will assume $\Sigma$ to be a 3-dimensional timelike hypersurface for our current interest. To describe the gravitational fields of domain walls, the metric off the walls satisfies vacuum Einstein's field equations with $\Lambda$
\bea
G_\c =   2 \Lambda * e_\c, \label{Einstein-eq}
\eea where $G_\c = R_{\a\b}\w * (e^\a \w e^\b \w e_\c)$ are Einstein's 3-forms and $R_{\a\b}$ are curvature 2-forms defined in terms of Levi-Civita connection $\nabla$ \cite{BT-87}. $e^\a$ are orthonormal co-frames and $*$ denotes the Hodge map associated with $g$. Moreover, by introducing the intrinsic metric $\hat{h}$ of $\Sigma$ \footnote{In the following, we will put $\hat{ }\,$ on any quantity to restrict it on $\Sigma$. }
\bea
\hat{h}= \hat{g} - \tilde{n} \otimes \tilde{n},
\eea
where $\tilde{n}= g(n, -)$ is the metric dual of unit normal $n$ of $\Sigma$, and also the extrinsic curvature $\pi_{ab}$ of $\Sigma$ defined by
\bea
\hat{\pi}_{ab}  = \frac{1}{2}(\mathcal{L}_n \bar{h} )_{ab}|_\Sigma, \label{pi}
\eea where $\mathcal{L}_n$ denotes the Lie derivative along $n$, and $\bar{h}$ is any extension of $h$ to a neighborhood of $\Sigma$, the metric on the $\Sigma$ should satisfy metric continuities, i.e. $g_{+}|_{\Sigma}=g_{-}|_{\Sigma}= \hat{g}$, and Israel's junction conditions
\bea
 \hat{\pi}_{ab+} - \hat{\pi}_{ab-} = -\frac{\kappa \sigma}{2} \,\hat{h}_{ab}, \label{IJC}
\eea where $\kappa= 8 \pi G$ and $\sigma=\textrm{constant}$ is the surface tension of domain walls.

In the spherical, planar, and hyperbolic symmetric space-time, the most general metric form can be written in double null-coordinates $(u, v)$ as
\bea
g= e^{2\mu (u, v)}   (- \d u \otimes \d v)  + B^2 (u, v) \d V_2, \label{g-1}
\eea where $\d V_2 = (1-k x^2)^{-1} \d x \otimes \d x +x^2 \d \phi \otimes \d \phi$, and constant Gaussian curvature $k= 1, 0, -1$ corresponds to 2-dimensional space-like spheres, planes, and hyperboloids, respectively.  In \cite{WCW-11}, a general non-degenerate solution of Eq. (\ref{Einstein-eq}) is obtained
\bea
g= 4 F(v) G(u) L(B) \,\d u \otimes \d v + B^2 \d V_2, \label{g-2}
\eea with $B(u, v)$ satisfies
\bea
\d B= - L(B) \,\,( F(v) \,\d v + G(u) \,\d u), \label{g-3}
\eea where $L(B)\equiv k  - \frac{2 M}{B} - \frac{\Lambda}{3} B^2$. It is clear that two arbitrary functions $F(v)$ and $G(u)$ are due to the freedom of choosing double-null coordinates. Eq. (\ref{g-3}) can be integrated to get
\bea
\mathcal{B}(B) = \mathcal{F}(v) + \mathcal{G}(u),
\eea where $\mathcal{B}(B)\equiv - \int L^{-1}\d B$, $\mathcal{F}(v) \equiv \int F\, \d v$, and $\mathcal{G}(u)\equiv \int G\, \d u$. $\mathcal{B}(B)$ in some particular choices of parameters $k$, $M$, $\Lambda$ are presented in Appendix \ref{A}.  If the inverse function $\mathcal{B}^{-1}$ of $\mathcal{B}(B)$ exists, which may only true in certain range of $B$,  we then obtain $B(u, v)=\mathcal{B}^{-1}(\mathcal{F} + \mathcal{G})$.

To obtain the domain wall solutions in comoving coordinates, it is convenient to introduce coordinate transformations $u=r+ \eta$ and $v=\frac{1}{r-\eta}$, and the wall is placed at $r=r_0$.  So metric $g_{+}$ and $g_{-}$, which correspond to $r>r_0$ and $r<r_0$ regions, give
\bea
g_{\pm}= A_{\pm}( \d \eta \otimes \d\eta - \d r\otimes \d r)  + B_{\pm}^2 \d V_2,
\eea  with
\bea
 \d B_{\pm} = -  L_{\pm}[(\frac{F_{\pm}}{(r-\eta)^2} + G_{\pm}) \d \eta
  + (\frac{- F_{\pm}}{(r-\eta)^2} + G_{\pm}) \d r]. \nn
\eea where $A_{\pm}\equiv \frac{4 F_{\pm}\, G_{\pm} \,L_{\pm}}{(r-\eta)^2}$ and $L_{\pm} \equiv (k-\frac{2 M_\pm}{B_{\pm}}-\frac{\Lambda_{\pm}}{3}B_{\pm}^2)$. The Israel's junction conditions give us two equations
\bea
&&  \zeta_1\,\frac{1}{\sqrt{-\hat{A}_+}}\widehat{\partial_r A_{+}}\, - \zeta_2\,\frac{1}{\sqrt{-\hat{A}_-}}\widehat{\partial_r A_{-}} =- \kappa\sigma \hat{A},\label{IJC-2}\\
&&  \zeta_1 \,\frac{\hat{B}_+\,\widehat{\partial_r B_{+}}}{\sqrt{-\hat{A}_+}} \, - \zeta_2\,\frac{\hat{B}_-\,\widehat{\partial_r B_{-}}}{\sqrt{-\hat{A}_-}} = -\frac{1}{2}\kappa\sigma \hat{B}^2 ,\label{IJC-3}
\eea where $\hat{A}$ and $\hat{B}$ denotes the metric components on $\Sigma$.  Here, $\{\zeta_1, \zeta_2\}= \pm 1$ due to the sign ambiguity of unit normal $n$ \cite{CGS-93}. In the following, we will only consider $\zeta_1= \zeta_2=1$.  Beside Israel's junction conditions, the metric continuities  also give two equations
\bea
\dot{\hat{\mathcal{F}}}_+\,\dot{\hat{\mathcal{G}}}_+\, \hat{L}_+=\dot{\hat{\mathcal{F}}}_-\,\dot{\hat{\mathcal{G}}}_-\hat{L}_-= \hat{A}/4, \label{MC-1} \\
\mathcal{B}_+^{-1}(\hat{\mathcal{F}}_+ + \hat{\mathcal{G}}_+) = \mathcal{B}_-^{-1}(\hat{\mathcal{F}}_- + \hat{\mathcal{G}}_-) = \hat{B}, \label{MC-2}
\eea where
 \bea \hat{\dot{\mathcal{F}}}_\pm \equiv \widehat{\partial_\eta \mathcal{F}_\pm}  = \frac{\hat{F}_\pm }{(r_0-\eta)^2}= \dot{\hat{\mathcal{F}}}_\pm, \\
 \hat{\dot{\mathcal{G}}}_\pm \equiv \widehat{\partial_\eta \mathcal{G}_\pm}  = \hat{G}_\pm= \dot{\hat{\mathcal{G}}}_\pm.
\eea

It can be showed that Eq. (\ref{IJC-2}) is implied by Eq. (\ref{IJC-3}). By differentiating Eq. (\ref{IJC-3}) with respect to $\eta$ and using Eqs (\ref{IJC-3})-(\ref{MC-2}), one can obtain Eq. (\ref{IJC-2}). So we now have three independent equations (\ref{IJC-3})-(\ref{MC-2}) for four unknown functions, $\hat{\mathcal{F}}_\pm$ and $\hat{\mathcal{G}}_\pm$. Before we discuss these equations, it is useful to define
\bea
&&R_{\pm}(u, v)= \mathcal{F}_\pm(v) + \mathcal{G}_\pm(u), \label{R}\\
&&T_{\pm}(u, v) =  \mathcal{F}_\pm(v) - \mathcal{G}_\pm(u), \label{T}
\eea so Eqs. (\ref{IJC-3})-(\ref{MC-2}) become
\bea
\hat{L}_+ \dot{\hat{T}}_+ - \hat{L}_- \dot{\hat{T}}_- =-\frac{\kappa\sigma}{2} \hat{B} \sqrt{-\hat{A}}\label{IJC-4}\\
(\dot{\hat{R}}_+^{\,2} - \dot{\hat{T}}_+^{\,2})\, \hat{L}_+ = (\dot{\hat{R}}_-^{\,2} - \dot{\hat{T}}_-^{\,2})\, \hat{L}_-= \hat{A},\label{MC-3}\\
 \mathcal{B}_+^{-1}(\hat{R}_+) = \mathcal{B}_-^{-1}(\hat{R}_-) = \hat{B}.\label{MC-4}
\eea By using Eqs. (\ref{MC-3})-(\ref{MC-4}), one can express $\dot{\hat{T}}_\pm$ in terms of $\dot{\hat{B}}$ and $\hat{A}$ as
\bea
\dot{\hat{T}}_{\pm} = h_{\pm} \sqrt{(-\hat{A} \hat{L}_\pm + \dot{\hat{B}}^{\,2})\hat{L}_\pm^{-2}}, \label{MC-5}
\eea where $\{h_+, h_-\}=\pm 1$ denotes the sign ambiguity coming from the quadratic in $\dot{\hat{T}}_\pm^{\,2}$.  It is clear that  Eq. (\ref{IJC-4}) will be used to determine $\hat{B}$, so $\hat{A}$ becomes a free function. The free choice of $\hat{A}$ comes from the freedom of choosing time coordinate on $\Sigma$. The different choices of $\hat{A}$ correspond to different time parametrization on $\Sigma$.

In the following, we will consider two different time parametrization, which are $\hat{A}=-1$ (case I) and $\hat{A}=-\alpha^2 / \eta^2$ (case II). We first consider case I, where the coordinate time $\eta$ on $\Sigma$ corresponds to proper time. So Eq. (\ref{IJC-4}) becomes
\bea
h_- \sqrt{(\hat{L}_- + \dot{\hat{B}}^{\,2})}- h_+ \sqrt{(\hat{L}_+ + \dot{\hat{B}}^{\,2})} =\frac{\kappa\sigma}{2} \hat{B}, \label{IJC-5}
\eea which is a well-known equation of motion of domain walls in the MW approach \cite{BKT-87, AKMS-87}. The intrinsic metric $\hat{g}$ obtained in CC approach are the same as in MW approach, so it is clear that these two approaches are equivalent. Eq. (\ref{IJC-5}) can also be written as
\bea
 \dot{\hat{B}}^2 = \frac{(\hat{L}_+ -\hat{L}_-)^2}{\kappa^2\sigma^2 \hat{B}^2} - \frac{1}{2} (\hat{L}_+ +\hat{L}_-) + \frac{\kappa^2\sigma^2}{16} \hat{B}^2, \label{IJC-6}
\eea and by substituting the definition of $\hat{L}_\pm$ into Eq. (\ref{IJC-6}) gives
\bea
 \dot{\hat{B}}^2 &=&  H^2 \hat{B}^2 - k  - \frac{\Delta M}{\hat{B}}\Big(\frac{4  (\Lambda_- -\Lambda_+)}{ 3 \kappa^2\sigma^2} -1+ \frac{2 M_-}{\Delta M}\Big) \nn\\
 &&+ \frac{4 (\Delta M)^2}{\kappa^2\sigma^2 \hat{B}^4}, \label{IJC-7}
\eea where $\Delta M\equiv M_+ - M_-$ and
\bea
H^2 \equiv \frac{\kappa^2\sigma^2 }{16} + \frac{(\Lambda_- -\Lambda_+)^2}{9 \kappa^2\sigma^2}  + \frac{(\Lambda_- +\Lambda_+) }{6}. \label{H}
\eea Eq. (\ref{IJC-7}) has been largely studied in the dynamics of bubbles and various exact solutions in some particular choices of parameters $M_\pm$ and $\Lambda_\pm$ have been obtained \cite{BKT-87, AKMS-87} (see also \cite{GV-89} and  references in).

In our previous work \cite{WCW-11},  a planar domain wall solution in de-Sitter space-time with refection symmetry is obtained, which satisfying the choice of $\hat{A}=-\alpha^2 / \eta^2$, and $\sigma_0$ vanishing gives the well-known metric of steady-state Universe in conformal time. We further use this domain solution to study gravitational effects of planar domain walls on primordial quantum fluctuations \cite{WWH-12}. In order to generalize our previous planar domain wall solutions, we study  spherical, planar, and hyperbolic domain wall space-times without reflection symmetry in case II. So Eq. (\ref{IJC-4}) becomes
\bea
h_- \sqrt{(\frac{\alpha^2 \hat{L}_-}{\eta^2} + \dot{\hat{B}}^{\,2})}- h_+ \sqrt{(\frac{\alpha^2 \hat{L}_+}{\eta^2}+ \dot{\hat{B}}^{\,2})} =\frac{\kappa\sigma}{2} \frac{\alpha \hat{B}}{\eta} \label{IJC-8}
\eea and substituting the definition of $\hat{L}_\pm$ into Eq. (\ref{IJC-8}) yields
\bea
\dot{\hat{B}}^2 &=& \frac{\alpha^2}{\eta^2} \Big[H^2 \hat{B}^2 - k  - \frac{\Delta M}{\hat{B}}\Big(\frac{4  (\Lambda_- -\Lambda_+)}{ 3 \kappa^2\sigma^2} -1+ \frac{2 M_-}{\Delta M}\Big) \nn\\
 &&+ \frac{4 (\Delta M)^2}{\kappa^2\sigma^2 \hat{B}^4}\Big]. \label{IJC-9}
\eea

Since solving Eqs. (\ref{IJC-7}) or (\ref{IJC-9}) only gives the metric $\hat{g}$ on $\Sigma$,  we should now discuss how to obtain the 4-dimensional metric $g_{\pm}$ in the comoving coordinates for given an exact solution of $\hat{B}$.  Suppose the two functions $\hat{A}(\eta)$ and $\hat{B}(\eta)$ are known, one can use Eqs. (\ref{MC-4}) and (\ref{MC-5}) to get $\hat{R}_{\pm}$ and $\hat{T}_{\pm}$. From Eqs. (\ref{R}) and (\ref{T}), we then obtain $\hat{\mathcal{F}}_\pm $ and $\hat{\mathcal{G}}_\pm $. By noting that $\mathcal{F}_{\pm}(v)$ are functions of $\frac{1}{r-\eta}$ and $\mathcal{G}_{\pm}(u)$ are functions of $r + \eta$, we learn that $\hat{\mathcal{F}}_\pm (\eta, r_0)$ and $\hat{\mathcal{G}}_\pm (\eta, r_0)$ are sufficient to give  $\mathcal{F}_{\pm}(v)$ and  $\mathcal{G}_{\pm}(u)$. It is easily to see that $A_\pm(r, \eta)$ and $B_\pm (r, \eta)$ can be derived from $\mathcal{F}_{\pm}(v)$ and  $\mathcal{G}_{\pm}(u)$, so we obtain domain wall solutions in the comoving coordinates.  In Sec. \ref{3}, we derive $A_\pm(r, \eta)$ and $B_\pm (r, \eta)$ in the special cases of $M_\pm=0$.



\section{Domain wall solutions in comoving coordinates} \label{3}

In this section we study domain wall solutions with $M_\pm=0$ in the comoving coordinates by choosing $\hat{g}_{00}=-1$ (case I)  and $\hat{g}_{00}=-\alpha^2/\eta^2$ (case II). The different choices of $\hat{g}_{00}$ correspond to different boundary conditions of $A_\pm(r, \eta)$ and $B_\pm (r, \eta)$.  In case I, we find that $A_\pm(r, \eta)$ and $B_\pm (r, \eta)$ are the same as the domain wall solutions obtained in \cite{CGS-93}. In case II, we only concentrate on $\Lambda_{\pm}>0$ and generalize our previous planar domain wall solutions to spherical, planar and hyperbolic domain walls in de Sitter space-time without reflection symmetry. When $M_\pm=0$, Eq. (\ref{IJC-4}) becomes
\bea
\dot{\hat{B}}^2 = \hat{A} (k- H^2\hat{B}^2), \label{IJC-10}
\eea and its exact solutions in the choice of $\hat{A}=-1$ have been studied in \cite{BKT-87, AKMS-87, GV-89}.

\subsection{Case I: ${\hat{A}=-1}$ and $M_\pm=0$}

Since Eq. (\ref{IJC-10}) has degenerate solutions in the case of $H=0$, we shall discuss $H\neq0$ and $H=0$ separately. In \cite{CGS-93}, Cveti\v{c} {\it et al} classified the domain wall solutions into extreme walls ($q_0=0$),  non- and ultraextreme walls ($q_0=\beta^2$) by the parameter $q_0$. Actually, it can be showed that the parameter $H^2$ corresponds to $q_0$ by rewritting Eq. (\ref{H}) to
\bea
\kappa\sigma = \pm\, 2\sqrt{H^2 -\frac{\Lambda_+}{3}} \mp 2 \sqrt{H^2 -\frac{\Lambda_-}{3}}, \label{H-2}
\eea which is the same as Eq. (2.34) in \cite{CGS-93}.

\subsubsection{$H^2\neq 0$} \label{H^2}
The exact solutions of Eq. (\ref{IJC-10}) in $k=\{-1, 0, 1\}$ yield
\bea
\hat{B}=
\left\{
\begin{array}{ll}
    \hat{B}^- =\frac{1}{H}\sinh H\eta, \label{B} \\
   \hat{B}^0 =e^{H\eta},\\
   \hat{B}^+ =\frac{1}{H}\cosh H\eta, 
  \end{array} \right.
 \eea where the superscripts $\{-, 0, +\}$ on $B$ and also any object in the following refer to $k=\{-1, 0, 1\}$, respectively. Moreover, Eq. (\ref{MC-4}) gives
 \bea
 &&\hat{R}^-_\pm=
\left\{
  \begin{array}{ll}
    {\lambda^{-1}_\pm} \tan^{-1} (\lambda_\pm \hat{B}^-), & \Lambda_\pm=3 \lambda^2_\pm, \label{R-1}\\
     \frac{1}{H}\sinh H\eta, & \Lambda_\pm=0,\\
    {\lambda^{-1}_\pm} \coth^{-1} ({\lambda_\pm} \hat{B}^-), & \Lambda_\pm=-3 \lambda^2_\pm,
  \end{array}
\right.  \\
&& \hat{R}^0_\pm = \left\{
  \begin{array}{ll}
   -\lambda^{-2}_\pm (\hat{B}^0)^{-1}, & \Lambda_\pm=3 \lambda^2_\pm, \\
    \,\,\,\,\lambda^{-2}_\pm  (\hat{B}^0)^{-1}, & \Lambda_\pm=-3 \lambda^2_\pm,
  \end{array}
\right.  \\
&&\hat{R}^+_\pm=
\left\{
  \begin{array}{ll}
   -   {\lambda^{-1}_\pm} \coth^{-1} ({\lambda_\pm} \hat{B}^+), & \Lambda_\pm=3 \lambda^2_\pm,
   \label{R-3}\\
  - \frac{1}{H}\cosh H\eta, & \Lambda_\pm=0,\\
    -  {\lambda^{-1}_\pm} \tan^{-1}  ({\lambda_\pm} \hat{B}^+), &\Lambda_\pm=-3 \lambda^2_\pm,

  \end{array}
\right.
 \eea where the inverse hyperbolic function $\coth^{-1}(\lambda_\pm \hat{B})$  in Eqs. (\ref{R-1}) and (\ref{R-3}) is obtained by considering $\hat{B}>\lambda^{-1}_\pm$. In the case of $\hat{B}<\lambda^{-1}_\pm$, we should obtain $\tanh^{-1}(\lambda_\pm \hat{B})$.\footnote{In the situation of  $\hat{B}<\lambda^{-1}_\pm$, one should use $\tanh^{-1}(\lambda_\pm \hat{B})$ and it turns out that the domain wall solutions $A_\pm(r, \eta)$ and $B_\pm(r, \eta)$ also yield Eqs. (\ref{B-})-(\ref{A+}). So these solutions are valid for both $\hat{B}>\lambda^{-1}_\pm$ and $\hat{B}<\lambda^{-1}_\pm$} By solving Eq. (\ref{MC-3}) yields
 \bea
 &&\hat{T}^-_\pm=
\left\{
  \begin{array}{ll}
   h_\pm {\lambda^{-1}_\pm} \tan^{-1} (\lambda_\pm \beta^-), & \Lambda_\pm=3 \lambda^2_\pm, \\
    h_\pm \frac{1}{H}\cosh H\eta, & \Lambda_\pm=0,\\
    -h_\pm {\lambda^{-1}_\pm} \coth^{-1} ({\lambda_\pm} \beta^-), & \Lambda_\pm=-3 \lambda^2_\pm,
  \end{array}
\right.  \\
&& \hat{T}^0_\pm = \left\{
  \begin{array}{ll}
   -  h_\pm \lambda^{-2}_\pm ({\beta}^0)^{-1}, & \Lambda_\pm=3 \lambda^2_\pm, \\
     h_\pm \lambda^{-2}_\pm  ({\beta}^0)^{-1}, & \Lambda_\pm=-3 \lambda^2_\pm,
  \end{array}
\right.  \\
&&\hat{T}^+_\pm=
\left\{
  \begin{array}{ll}
   -  h_\pm   {\lambda^{-1}_\pm} \coth^{-1} ({\lambda_\pm} \beta^+), & \Lambda_\pm=3 \lambda^2_\pm,\\
    h_\pm \frac{1}{H}\sinh H\eta, & \Lambda_\pm=0,\\
      h_\pm  {\lambda^{-1}_\pm} \tan^{-1}  ({\lambda_\pm} \beta^+), &\Lambda_\pm=-3 \lambda^2_\pm,\label{T-1}

  \end{array}
\right.
 \eea where
 \bea
 \left\{
\begin{array}{ll}
    \beta^- =\sqrt{(H^2- \Lambda_\pm/3)^{-1}}\cosh H\eta, \\
   \beta^0 =H\,\sqrt{(H^2-\Lambda_\pm/3)^{-1}}\,e^{H\eta},\\
  \beta^+ =\sqrt{(H^2-\Lambda_\pm/3)^{-1}}\sinh H\eta, 
  \end{array} \right.
 \eea with $H^2>\lambda_\pm^2$.\footnote{In the case of $H^2=\lambda^2_\pm$, $\hat{T}_\pm$ becomes constants and the solutions of $A_\pm(r, \eta)$ and $B_\pm(r, \eta)$ correspond to $\gamma_\pm=0$ in Eqs. (\ref{B-})-(\ref{A+}).} Substituting Eqs. (\ref{R-1})-(\ref{T-1}) into Eqs. (\ref{R})-(\ref{T}), and noting that $\mathcal{F}(v)$ and $\mathcal{G}(u)$ are functions of $1/(r-\eta)$ and $r + \eta$, respectively, we can obtain $\mathcal{F}(v)$ and $\mathcal{G}(u)$. Since the calculations are straightforward and similar for $\Lambda_\pm>0$ and $\Lambda_\pm <0$,  we will only present the results for $\Lambda_\pm\geqslant0$ with $h_\pm =1$, which yields
 \bea
 &&\mathcal{F}^-_\pm=
\left\{
  \begin{array}{ll}
    \frac{1}{2 \lambda_\pm} \tan^{-1} \{\frac{1}{\sinh(\gamma_\pm+ H (\frac{1}{v}- r_0))}\}, &\Lambda_\pm>0, \\
     \frac{1}{2 H} e^{-H(\frac{1}{v} - r_0)}, & \Lambda_\pm=0,\\
  \end{array}
\right.  \\
&& \mathcal{F}^0_\pm = \left\{
  \begin{array}{ll}
   - \frac{1}{2 H \lambda_\pm} e^{-(H r_0-\gamma_\pm)}\,e^{H/v}, & \Lambda_\pm>0, \\
  \end{array}
\right.  \\
&&\mathcal{F}^+_\pm=
\left\{
  \begin{array}{ll}
     \frac{- \coth^{-1} \{\cosh( \gamma_\pm + H(\frac{1}{v}- r_0))\}}{2\lambda_\pm}, & \Lambda_\pm>0, \\
  -  \frac{1}{2 H} e^{H(\frac{1}{v} - r_0)}, & \Lambda_\pm=0,\\

  \end{array}
\right.
 \eea and
 \bea
 &&\mathcal{G}^-_\pm=
\left\{
  \begin{array}{ll}
  -  \frac{1}{2 \lambda_\pm} \tan^{-1} \{\frac{1}{\sinh(\gamma_\pm+ H (u - r_0))}\}, &\Lambda_\pm>0, \\
   -  \frac{1}{2 H} e^{-H( u - r_0)}, & \Lambda_\pm=0,\\
  \end{array}
\right.  \\
&& \mathcal{G}^0_\pm = \left\{
  \begin{array}{ll}
   - \frac{1}{2 H \lambda_\pm} e^{(H r_0-{\gamma_\pm})}\,e^{-Hu}, & \Lambda_\pm>0, \\
  \end{array}
\right.  \\
&&\mathcal{G}^+_\pm=
\left\{
  \begin{array}{ll}
   -  \frac{\coth^{-1} \{\cosh( \gamma_\pm + H(u - r_0))\}}{2 \lambda_\pm}, & \Lambda_\pm>0, \\
   -\frac{1}{2 H} e^{H( u- r_0)}, & \Lambda_\pm=0,\\

  \end{array}
\right.
  \eea where
\bea
 \gamma_\pm=\ln\{\,\lambda_\pm^{-1}H + \sqrt{-1 +H^2/\lambda_\pm^2}\}=\cosh^{-1}(\sqrt{H^2/\lambda_\pm^2}).\nn
 \eea The formula
 $\tan^{-1}x \pm \tan^{-1}y = \tan^{-1}(\frac{x \pm y}{1\mp x y})$ and
$ \coth^{-1}x \pm \coth^{-1} y = \coth^{-1}(\frac{xy\pm 1}{y\pm x})$
 have been used to calculate $\mathcal{F}_\pm(v)$ and $\mathcal{G}_\pm(u)$.

 Since $\mathcal{F}_\pm(v)$ and $\mathcal{G}_\pm(u)$ are obtained, one can direct calculate $B_\pm(r, \eta)$ and $A_\pm(r, \eta)$ to get
 \bea
 && B^-_\pm=
\left\{
  \begin{array}{ll}
    \frac{\sinh H\eta}{\lambda_\pm\cosh\{\gamma_\pm +H(r-r_0)\}}, &\Lambda_\pm>0, \label{B-}\\
     \frac{1}{H} e^{-H(r - r_0)}\sinh H\eta, & \Lambda_\pm=0,\\
  \end{array}
\right.  \\
&& B^0_\pm = \left\{
  \begin{array}{ll}
    \frac{H e^{H\eta}}{\lambda_\pm \cosh\{H(r-r_0)+\gamma_\pm\}}, & \Lambda_\pm>0, \label{B0}\\
  \end{array}
\right.  \\
&&B^+_\pm=
\left\{
  \begin{array}{ll}
    \frac{\cosh H\eta}{\lambda_\pm\cosh\{\gamma_\pm + H(r-r_0)\}}, & \Lambda_\pm>0, \label{B+} \\
   \frac{1}{H} e^{H(r - r_0)}\cosh H\eta, & \Lambda_\pm=0,\\

  \end{array}
\right.
 \eea and
 \bea
 && A^-_\pm=
\left\{
  \begin{array}{ll}
    -\frac{H^2}{\lambda^2_\pm[\cosh\{\gamma_\pm +H(r-r_0)\}]^2}, &\Lambda_\pm>0, \\
     - e^{-2 H(r - r_0)}, & \Lambda_\pm=0,\\
  \end{array}
\right.  \\
&& A^0_\pm = \left\{
  \begin{array}{ll}
    -\frac{H^2}{\lambda^2_\pm [\cosh\{H(r-r_0)+\gamma_\pm\}]^2}, & \Lambda_\pm>0, \label{A0}\\
  \end{array}
\right.  \\
&&A^+_\pm=
\left\{
  \begin{array}{ll}
    - \frac{H^2}{\lambda^2_\pm[\cosh\{\gamma_\pm + H(r-r_0)\}]^2}, & \Lambda_\pm>0, \\ \label{A+}
    - e^{2 H(r - r_0)}, & \Lambda_\pm=0,\\

  \end{array}
\right.
\eea which agree with the results for $q_0=\beta^2$ in \cite{CGS-93}.\footnote{We should point out that the $\gamma_\pm$ appeared in hyperbolic cosine are different from the integration constant $\beta z''$ in \cite{CGS-93} due to different process of normalization.} It is not difficult to verify that the domain wall solutions for $\Lambda_\pm <0$ are also equivalent to Cveti\v{c} {\it et al}'s results \cite{CGS-93}. When $3 H^2=\Lambda_+= \Lambda_- > 0$, Eq. (\ref{H-2}) yields  $\sigma=0$, which means no domain wall in the space-times. So it corresponds to $\gamma_+=\gamma_-=0$ in the solutions (\ref{B-})-(\ref{A+}).

\subsubsection{H=0}

In the case of $H=0$, which corresponds to extreme walls in \cite{CGS-93}, the exact solutions of Eq. (\ref{IJC-10}) give
\bea
\hat{B}=
\left\{
\begin{array}{ll}
    \hat{B}^- =\eta, \\
   \hat{B}^0 =1,
  \end{array} \right.
 \eea where no real solution exists in $k=1$.  Since the following calculations to obtain $\mathcal{F}_\pm$ and $\mathcal{G}_\pm$ are similar to the calculations in the previous subsection \ref{H^2}, we directly present the final results of $\mathcal{F}_\pm$ and $\mathcal{G}_\pm$ with $h_\pm =1$, which yield
  \bea
 &&\mathcal{F}^-_\pm=
\left\{
  \begin{array}{ll}
    \frac{1}{2 \lambda_\pm}\ln\{ r_0 + \lambda_\pm-\frac{1}{v}\}, &\Lambda_\pm= - 3 \lambda_\pm^2, \\ \label{FF-1}
  \frac{1}{2}(b_\pm +  r_0 - \frac{1}{ v} ), & \Lambda_\pm=0,\\
  \end{array}
\right.  \\
&& \mathcal{F}^0_\pm = \left\{
  \begin{array}{ll}
   \frac{1}{2}(\lambda^{-2}_{\pm} + \lambda^{-1}_\pm r_0 - \frac{1}{\lambda_\pm v} ), & \Lambda_\pm= - 3 \lambda_\pm^2, \\
  \end{array}
  \right.
 \eea and
 \bea
 &&\mathcal{G}^-_\pm=
\left\{
  \begin{array}{ll}
 - \frac{1}{2 \lambda_\pm}\ln\{ u- r_0 - \lambda_\pm\}, &\Lambda_\pm= - 3 \lambda_\pm^2, \\
   \frac{1}{2}(-b_\pm - r_0 + u ),  & \Lambda_\pm=0,\\
  \end{array}
\right.  \\
&& \mathcal{G}^0_\pm = \left\{
  \begin{array}{ll}
     \frac{1}{2}(\lambda^{-2}_{\pm} + \lambda^{-1}_\pm r_0 - \frac{u}{\lambda_\pm} ),  & \Lambda_\pm= - 3 \lambda_\pm^2,  \label{GG-1}\\
  \end{array}
\right.
   \eea where $b_\pm$ are constants and no real solution exists for $\Lambda_\pm >0$. From Eqs. (\ref{FF-1})-(\ref{GG-1}), one can directly calculate $B_\pm(r, \eta)$ and $A_\pm(r, \eta)$ to obtain
  \bea
 &&B^-_\pm=
\left\{
  \begin{array}{ll}
   \frac{- \eta}{\lambda_\pm (r-r_0) -1}, &\Lambda_\pm= - 3 \lambda_\pm^2, \\ \label{FF-1}
  \eta, & \Lambda_\pm=0,\\
  \end{array}
\right.  \\
&& B^0_\pm = \left\{
  \begin{array}{ll}
  \frac{-1}{\lambda_\pm (r-r_0) - 1}, & \Lambda_\pm= - 3 \lambda_\pm^2, \\
  \end{array}
  \right.
 \eea and
 \bea
 &&{A}^-_\pm=
\left\{
  \begin{array}{ll}
 - \frac{1}{(\lambda_\pm (r-r_0) -1)^2}, &\Lambda_\pm= - 3 \lambda_\pm^2, \\
  -1,  & \Lambda_\pm=0,\\
  \end{array}
\right.  \\
&& {A}^0_\pm = \left\{
  \begin{array}{ll}
    - \frac{1}{(\lambda_\pm (r-r_0) - 1)^2}.  & \Lambda_\pm= - 3 \lambda_\pm^2,  \label{GG-1}\\
  \end{array}
\right.
   \eea which corresponds to extreme wall solutions $(q_0=0)$ in \cite{CGS-93}. So domain wall solutions in case I yield the same solutions as in \cite{CGS-93}. It indicates that one can transform domain wall solutions in static background metric to Cveti\v{c} {\it et al}'s domain wall solutions \cite{CGS-93} by coordinate transformations.

 \subsection{Case II: $\hat{A}=-\alpha^2/\eta^2$ and $M_\pm=0$}

In our previous work \cite{WCW-11}, we obtained a planar domain wall solution with reflection symmetry in de Sitter space-time, where $\hat{A}=-\alpha^2/\eta^2 $, and then study its gravitational effects on primordial quantum fluctuations during inflation \cite{WWH-12}. In case II, we generalize our previous planar domain wall solution to spherical, planar, and hyperbolic domain walls without reflection symmetry. In the choice of $\hat{A}=-\alpha^2/\eta^2 $, Eq. (\ref{IJC-10}) becomes
\bea
\dot{\hat{B}}^2 = - \frac{\alpha^2}{\eta^2} \,(k- H^2\hat{B}^2). \label{IJC-12}
\eea and exact solutions of Eq. (\ref{IJC-12}) yield
\bea
\hat{B}=
\left\{
\begin{array}{ll}
    \hat{B}^- =\frac{\eta}{2H} - \frac{1}{2\eta H}, \\  \label{B-2}
   \hat{B}^0 =-\frac{1}{\eta},\\
   \hat{B}^+ =\frac{\eta}{2H} + \frac{1}{2\eta H}, 
  \end{array} \right.
\eea where we have set $\alpha^2=1/H^2$. One may notice that the coordinate time $\eta$ is related to proper time $\tau$  on $\Sigma$ by $\eta= -e^{-H \tau}$, where Eq. (\ref{B-2}) becomes Eq. (\ref{B}) (up to a sign choice). So $\eta$ may be considered as conformal time in de Sitter space-time. In the following, we only discuss $\Lambda_\pm >0$ and $H^2 \neq 0$.

From Eq. (\ref{MC-4}) we obtain
\bea
\hat{R}=
\left\{
\begin{array}{ll}
    \hat{R}_\pm^- =  {\lambda^{-1}_\pm} \tan^{-1} (\lambda_\pm \hat{B}^-), \\
   \hat{R}_\pm^0 = -\lambda^{-2}_\pm (\hat{B}^0)^{-1},\\
   \hat{R}_\pm^+ = -   {\lambda^{-1}_\pm} \coth^{-1} ({\lambda_\pm} \hat{B}^+), 
  \end{array} \right.
\eea and then solving Eq. (\ref{MC-3}) gives
\bea
\hat{T}=
\left\{
\begin{array}{ll}
    \hat{T}_\pm^- =  -{\lambda^{-1}_\pm} \tan^{-1} \{\lambda_\pm\mu_\pm (\frac{\eta}{H}-\hat{B}^-)\}, \\
   \hat{T}_\pm^0 = h_\pm \lambda^{-2}_\pm (\mu_\pm\hat{B}^0)^{-1},\\
   \hat{T}_\pm^+ =   {\lambda^{-1}_\pm} \coth^{-1} \{\lambda_\pm\mu_\pm(\frac{\eta}{H}- \hat{B}^+\}, 
  \end{array} \right.
\eea where $\mu_\pm=\sqrt{(1-\lambda^2_\pm/H^2)^{-1}}$ and $h_\pm=1$ are chosen in the case of spherical and hyperbolic walls.

A tedious but straightforward calculation of $\mathcal{F}(v)$ and $\mathcal{G}(u)$ gives
\bea
\mathcal{F}_\pm=
\left\{
\begin{array}{ll}
   \mathcal{F}^-_\pm = \frac{1}{2\lambda_\pm}\tan^{-1}\{\frac{2}{  e^{\gamma_\pm} (\frac{1}{v}-r_0)- e^{-\gamma_\pm} (\frac{1}{v}-r_0)^{-1}  }\}, \\
  \mathcal{F}^0_\pm = -\frac{1}{2\lambda_\pm H} e^{\{h_\pm\gamma_\pm\}} (\frac{1}{v}-r_0),\\
  \mathcal{F}^+_\pm = \frac{1}{2\lambda_\pm}\coth^{-1}\{ \frac{e^{-\gamma_\pm} (\frac{1}{v}-r_0)^{-1}+  e^{\gamma_\pm} (\frac{1}{v}-r_0)}{2}\}, 
  \end{array} \right.
\eea and
 \bea
\mathcal{G}_\pm=
\left\{
\begin{array}{ll}
   \mathcal{G}^-_\pm =- \frac{1}{2\lambda_\pm}\tan^{-1}\{\frac{2}{  e^{-\gamma_\pm} (u-r_0)- e^{\gamma_\pm}( \frac{1}{u-r_0})  }\},  \\
  \mathcal{G}^0_\pm = \frac{1}{2\lambda_\pm H}\, e^{\{-h_\pm\gamma_\pm\}} (u-r_0),\\
  \mathcal{G}^+_\pm =-\frac{1}{2\lambda_\pm}\coth^{-1}\{ \frac{e^{\gamma_\pm} (\frac{1}{u-r_0})+  e^{-\gamma_\pm} (u-r_0)}{2}\}. 
  \end{array} \right.
\eea So one can then calculate $A(r, \eta)$ and $B(r, \eta)$ to get
 \bea
B_\pm=
\left\{
\begin{array}{ll}
   B^-_\pm = \frac{ \eta^2-(r-r_0)^2-1 }{2 \{ H\eta-\sqrt{H^2-\lambda_\pm^2}\,\,(r-r_0)\} }, \label{B-3}\\
  B^0_\pm = \frac{ -H }{  H \eta- h_\pm \sqrt{(H^2-\lambda^2_\pm)}\,\,(r-r_0) },\\
  B^+_\pm =\frac{ \eta^2-(r-r_0)^2+1 }{2 \{ H\eta-\sqrt{H^2-\lambda_\pm^2}\,\,(r-r_0)\} }, 
  \end{array} \right.
\eea and
 \bea
A_\pm=
\left\{
\begin{array}{ll}
   A^-_\pm =- \frac{ 1 }{ (H\eta-\sqrt{H^2-\lambda_\pm^2}\,\,(r-r_0))^2 },  \\  \label{A+2}
  A^0_\pm =- \frac{ 1 }{ ( H \eta- h_\pm \sqrt{(H^2-\lambda^2_\pm)}\,\,(r-r_0) )^2},\\
  A^+_\pm =-\frac{ 1 }{( H\eta-\sqrt{H^2-\lambda_\pm^2}\,\,(r-r_0))^2 }. 
  \end{array} \right.
\eea Eqs. (\ref{B-3}) and (\ref{A+2}) are the metric solutions of spherical, planar, hyperbolic domain walls in de Sitter space-times. In the case of  $h_-=-h_+=1$ and $\Lambda_-=\Lambda_+$, the solutions $B^0_\pm(r, \eta)$ and $A^0_\pm(r, \eta)$ do return to our previous planar domain wall solution \cite{WCW-11}. It is clear that planar domain wall space-times with different positive $\Lambda$ on each side of the wall are conformally flat, and when $\sigma=0$, which corresponds to $H^2= \lambda^2_+=\lambda^2_-$,  $B^0_\pm(r, \eta)$ and $A^0_\pm(r, \eta)$ become the metric for describing inflationary Universe in the conformal time coordinate \cite{BD-82, Linde-90}. Since quantum fluctuations during inflation \cite{LL-00, Linde-90, Dodelson-03} and the well-known definition of the vacuum state, i.e. Bunch-Davies vacuum \cite{BD-87}, are described in background metric  $B^0_\pm(r, \eta)$ and $A^0_\pm(r, \eta)$ with $\sigma=0$,  it suggests that  this co-moving coordinates in case II is a proper coordinate choice to study gravitational effects of domain walls during inflation. In \cite{WWH-12}, we study  quantum fluctuations of a scalar field in background planar domain wall metric, $B^0_\pm(r, \eta)$ and $A^0_\pm(r, \eta)$, with reflection symmetry, and it yields that gravitational effects of planar domain wall will cause a primordial dipole in the power spectrum of primordial curvature perturbations.  In the spherical domain wall solution, i.e. $B^+_\pm(r, \eta)$ and $A^+_\pm(r, \eta)$, one may define a local neighborhood $\mathcal{N}_p$ by restricting coordinates $\{\eta, r, x, \phi\}$ to the range $\eta^2-(r-r_0)^2\ll 1$ and $x\ll 1$, so gravitational fields of spherical domain walls in $\mathcal{N}_p$ can well approximate as the planar domain wall metric, i.e. $B^0_\pm(r, \eta)$ and $A^0_\pm(r, \eta)$.

The global structure of domain wall space-times in case I have been well studied in \cite{BKT-87, CGS-93, CS-97}. In Sec. \ref{4}, we will first present coordinate transformations between the domain wall solutions Eqs. (\ref{B-})-(\ref{A+}) in case I and the solutions Eqs. (\ref{B-3})-(\ref{A+2}) in case II, and then study the global structure of domain wall space-times of case II.

\section{Global structure of domain wall space-times} \label{4}

The global structure of spherical, planar and hyperbolic domain walls has been largely discussed in \cite{CGS-93, BKT-87} (see the review article \cite{CS-97} and the references in). Ref. \cite{CS-97} has pointed out that the constant-$r$ sections of domain wall solutions in case I with $H^2\neq0$, i.e. Eqs. (\ref{B-})-(\ref{A+}),
all represent (2+1)-dimensional de Sitter space-time ($\d\textrm{S}_3$), whose topology is $\textbf{R}$ ({time}) $\times$ $\textbf{S}^2$ ({space}).
It turns out that planar domain wall solution, i.e. Eqs. (\ref{B0}) and (\ref{A0}), which is locally plane-symmetric and geodesically incomplete, describes only a part of a spherical bubble \cite{CS-97}. Since the metric of spherical domain wall solution, Eqs. (\ref{B+}) and (\ref{A+}), internal to the wall is geodesically complete, Ref. \cite{CGS-93} study geodesic extension of spherically non- and ultraextreme domain wall space-time with non-positive $\Lambda_\pm$  in $(\eta, r)$ directions.

It is useful to study global and casual structure of  space-times by using conforaml diagram \cite{HE-73, Wald-84}, which compactify space-time infinity into finite region. To study the global structure of domain wall solution in case II, we first present the coordinate transformation between the  domain wall solutions in case I and case II. By performing the following coordinate transformations
\bea
\begin{array}{ll}
&\eta = - \cosh \{H(r'-r'_0)\} e^{-H t},\label{CT}\\ 
&r-r_0= \sinh \{H(r'-r'_0)\} e^{-Ht},
\end{array}
\eea on domain wall metric in case II, we then obtain $d s^2_\pm=$
\bea
\left\{
\begin{array}{ll}
     \frac{H^2}{\lambda^2_\pm[\cosh\{\gamma_\pm +H(r'-r'_0)\}]^2}(-\d t^2 + \d r'^2+ [\frac{\sinh H t}{H}]^2 \d H^2_2 ) \label{ds}\\
 \frac{H^2}{\lambda^2_\pm[\cosh\{\gamma_\pm +H(r'-r'_0)\}]^2}(-\d t^2 + \d r'^2+ e^{2 H t} \d X^2_2 ),\\
  \frac{H^2}{\lambda^2_\pm[\cosh\{\gamma_\pm +H(r'-r'_0)\}]^2}(-\d t^2 + \d r'^2+ [\frac{\cosh H t}{H}]^2 \d \Omega^2_2 ), 
  \end{array} \right.
\eea which correspond to non- and ultraextreme domain wall solutions in case I. Here, $\d H^2_2$, $\d X^2_2$ and $\d \Omega^2_2$ denote the line element of two-dimensional hyperboloid, plane, and sphere respectively. From the coordinate transformations (\ref{CT}), it is clear that both $\{\eta, r, x, \phi\}$ and $\{t, r', x, \phi\}$ are comoving coordinates with the wall sitting at $r=r_0$ and $r'=r'_0$, respectively. Moreover, Eq. (\ref{CT}) restrict the $\eta$-coordinate range to $\eta\leqslant0$, so the future infinity ($t=\infty$) and past infinity ($t=-\infty$) in the conformal diagram of case I domain wall solutions ($H^2\neq0$) corresponds to $\eta=0$ and $\eta=-\infty$. The geodesic extensions of the comoving coordinate patch have been present in \cite{CGS-93, CS-97}. By transforming the domain wall solutions in case II to Eq. (\ref{ds}), we realize that the topology of spherical, planar, and hyperbolic domain walls in case II is also $\textbf{R}$ ({time}) $\times$ $\textbf{S}^2$ ({space}).

\section{Discussion} \label{5}
We have present a proof on the equivalence of the CC approach and MW approach, and shown how to construct a comoving coordinate by knowing the trajectories of domain walls in the MW approach. The spherical, planar and hyperbolic domain wall solutions with $M_\pm=0$ are obtained in two different comoving coordinates, which are referred to case I and case II. The case I domain wall solutions yield the same solutions obtained in \cite{CGS-93}. Refs. \cite{CGS-93, CS-97} studied the global and casual structure of case I domain wall solutions. As pointed out in \cite{CGS-93}, one can observe that constant-$r$ sections of spherical, planar and hyperbolic domain wall solutions with $H^2\neq 0$ in case I represent  (2+1)-dimensional de Sitter space-time $\d\textrm{S}_3$. Since it is not clear to us whether spherical, plane and hyperbolic domain walls in case II also represent $\d\textrm{S}_3$, we find the coordinate transformation between case I and case II domain wall solutions with $H^2\neq0$. Hence, one can understand that the planar domain wall solution in case II represents a portion of a spherical domain wall space-time. From the coordinate transformation Eq. (\ref{CT}), we also learn the future infinity ($t=\infty$) and past infinity ($t=-\infty$) in the conformal diagram of case I domain wall solutions ($H^2\neq0$) corresponds to $\eta=0$ and $\eta=-\infty$ in the case II domain wall solutions.

In \cite{WCW-11},  we obtain a planar domain wall solution, which is conformally flat. When $\sigma=0$, the planar domain wall metric returns to the well-known metric of steady state Universe, which has been used to study quantum fluctuations during inflation, in conformal time coordinate. So we studied quantum fluctuations of a scalar field in the planar domain wall space-time and find the gravitational effects of the planar domain wall on power spectrum of primordial curvature perturbation \cite{WWH-12}. However, the planar domain wall solution requires plane and reflection symmetry, which is quite limited in the study of gravitational fields of realistic domain wall space-times. For example, false vacuum decay yields two space-time regions with different $\Lambda$ separated by a spherical bubble. So the case II domain wall solutions, i.e. Eqs. (\ref{B-3})-(\ref{A+2}), generalized our previous planar domain wall solution \cite{WCW-11} to spherical, planar and hyperbolic domain walls without refection symmetry. These solutions will be useful for further investigation on primordial quantum fluctuation of scalar fields during inflation.

In this paper, we only discuss domain wall solutions in the case of $M_\pm=0$. There are various interesting physical problems, which may need to consider $M_\pm\neq0$. For example, in the study the evolution of remnants of the false vacuum surrounded by the true vacuum \cite{BKT-87}, one should consider $M_+\neq0$.  So finding domain wall solutions in the comoving coordinates with $M_\pm\neq0$ may also be useful for studying some physical problems and their associated global structure of space-times. This will be considered as our future works.
\acknowledgments

CHW is supported by the National Science Council of the Republic of China under the grants NSC 98-2811-M-032-005 and YHW is fully supported by the NCU Top University Project funded by the Ministry of Education, Taiwan ROC and Center for Mathematics and Theoretical Physics, National Central University. This work is also partially support by the National Center for Theoretical Sciences (NCTS), Hsinchu.

\begin{appendix}
\section{Integral form of $\mathcal{B}(B)$} \label{A}

Since the calculations of domain wall solutions in Sec. \ref{2} and Sec. \ref{3} involve integral function $\mathcal{B}$, which is
\bea
\mathcal{B}(B_\pm)= \int \frac{d B_\pm}{-k + \frac{2M_\pm}{B_\pm} + \frac{\Lambda_\pm}{3} B_\pm^2}, \label{A-1}
\eea  we present the integration of Eq. (\ref{A-1}) here. In the following, we omit the subscript $\pm$ for simplification. To integrate Eq. (\ref{A-1}) in the case of $k=1$ and $\{M, \Lambda\}>0$, we should first study the cubic equations $B^3 - \frac{3}{\Lambda}  B + \frac{6M}{\Lambda}=0$, and if determinant  $D= -\frac{1}{\Lambda^3} + \frac{9 M^2}{\Lambda^2}<0$, there are three distinct real roots (2 positive and 1 negative roots)
\bea
 b_1 = 2 \sqrt{\frac{1}{\Lambda}}\cos{\frac{\phi}{3}}, \hspace{0.5cm}
     b_{2,3} = -2 \sqrt{\frac{1}{\Lambda}}\cos{\frac{\phi\pm\pi}{3}},
\eea where $\phi =  \arccos\{ -{3 M}\sqrt{\Lambda}\} $. Integrating Eq. (\ref{A-1}) yields
\bea
\mathcal{B}(B)=
\left\{
\begin{array}{ll}
 \frac{3\ln\{ (B- b_{++})^\mu(B-b_+)^\nu (B-b_-)^\rho\}}{\Lambda},  B> b_{++}, \label{A-2} \\
  \frac{3\ln\{ (b_{++}- B)^\mu(B-b_+)^\nu (B-b_-)^\rho\}}{\Lambda},  b_{++}>B>b_+,\\
 \frac{3\ln\{ ( b_{++}- B)^\mu(b_+- B)^\nu (B-b_-)^\rho\}}{\Lambda},   b_{+}>B,  
  \end{array} \right.
\eea with $\mu$, $\nu$ and $\rho$ satisfying
\bea
\frac{\mu}{(B-b_{++})} + \frac{\nu}{(B-b_{+})} + \frac{\rho}{(B-b_{-})} = \frac{B}{B^3 - \frac{3}{\Lambda}  B + \frac{6M}{\Lambda}},\nn
\eea where $b_{++}$ denotes the larger positive root and $b_-$ denotes the negative root. From Eq. (\ref{A-2}), we learn that one coordinate chart can only cover either $B>b_+$ or $B<b_+$, so one need to introduce separate coordinate charts to cover the whole range of $B$.

In the case of $M=0$ and $k=1$, the integration of Eq. (\ref{A-1}) gives
\bea
\mathcal{B}(B)=
\left\{
\begin{array}{ll}
-\lambda^{-1}\coth^{-1}(\lambda B), & B>\lambda^{-1},\\ \label{A-3}
-\lambda^{-1}\tanh^{-1}(\lambda B), & B< \lambda^{-1},
\end{array} \right.
\eea for $\Lambda= 3\lambda^2$, and
\bea
\mathcal{B}(B)=
-\lambda^{-1}\tan^{-1}(\lambda B),
\eea for $\Lambda= - 3\lambda^2$. Similarly, in the case of $M=0$ and $k=-1$, we obtain
\bea
\mathcal{B}(B)=
\lambda^{-1}\tan^{-1}(\lambda B),
\eea for $\Lambda= 3\lambda^2$ and
\bea
\mathcal{B}(B)=
\left\{
\begin{array}{ll}
\lambda^{-1}\coth^{-1}(\lambda B), & B>\lambda^{-1},\\
\lambda^{-1}\tanh^{-1}(\lambda B), & B< \lambda^{-1},
\end{array} \right.
\eea for $\Lambda= -3\lambda^2$. Finally, the case of $M=k=0$ yields
\bea
\mathcal{B}(B)= -\frac{3}{\Lambda} B^{-1}. \label{A-4}
\eea Eqs. (\ref{A-3})-(\ref{A-4}) will be used to calculate  $A_\pm(r, \eta)$ and $B_\pm(r, \eta)$ in Sec. \ref{3}, and it is not difficult to verify that $A_\pm$ and $B_\pm$ yield the same results in the calculation of $B>\lambda^{-1}$ and $B<\lambda^{-1}$

\end{appendix}

\end{document}